\documentclass[useAMS, usenatbib]{biom}

\usepackage{amsfonts} 
\usepackage{mathtools} 
\usepackage{multirow} 
\usepackage{amssymb} 

\newcommand{\bci}{$\bullet$}
\newcommand{\wci}{$\circ$}
\newcommand{\bwci}{\rlap{\bci}\,\,\,\wci}
\newcommand{\wbci}{\rlap{\wci}\,\,\,\bci}

\newcommand{\bsq}{{\tiny$\blacksquare$}}
\newcommand{\wsq}{{\tiny$\square$}}
\newcommand{\bwsq}{\rlap{\bsq}\,\,\,\wsq}
\newcommand{\wbsq}{\rlap{\wsq}\,\,\,\bsq}

\newcommand{\btr}{{\scriptsize$\blacklozenge$}}
\newcommand{\wtr}{{\scriptsize$\lozenge$}}
\newcommand{\bwtr}{\rlap{\btr}\,\,\,\wtr}
\newcommand{\wbtr}{\rlap{\wtr}\,\,\,\btr}

\title[Cumulative Link Mixed-Effects Models in the Service of Remote Sensing Crop Progress Monitoring]{Cumulative Link Mixed-Effects Models in the Service of Remote Sensing Crop Progress Monitoring}

\author{Ioannis Oikonomidis$^{*}$\email{goikon@math.uoa.gr} and
Samis Trevezas$^{**}$\email{strevezas@math.uoa.gr} \\
Department of Mathematics, National and Kapodistrian University of Athens, Athens, Greece}

\begin{document}
%
%
\label{firstpage}
\begin{abstract}
This study introduces an innovative Cumulative Link Modeling approach to monitor crop progress over large areas using remote sensing data. The models utilize the predictive attributes of calendar time, thermal time, and the Normalized Difference Vegetation Index (NDVI). Two distinct issues are tackled: real-time crop progress prediction, and completed season fitting. In the context of prediction, the study presents two model variations, the standard one based on the Multinomial distribution and a novel one based on the Multivariate Binomial distribution. In the context of fitting, random effects are incorporated to capture the inherent inter-seasonal variability, allowing the estimation of biological parameters that govern crop development and determine stage completion requirements. Theoretical properties in terms of consistency, asymptotic normality, and distribution-misspecification are reviewed. Model performance was evaluated on eight crops, namely corn, oats, sorghum, soybeans, winter wheat, alfalfa, dry beans, and millet, using in-situ data from Nebraska, USA, spanning a 20-year period. The results demonstrate the wide applicability of this approach to different crops, providing real-time predictions of crop progress worldwide, solely utilizing open-access data. To facilitate implementation, an ecosystem of R packages has been developed and made publicly accessible under the name Ages of Man.
\end{abstract}
\maketitle
\section{Introduction}
\label{s:introduction}
The agricultural sector has been experiencing intensifying automation in recent decades. Against the pressure posed by the ever-increasing human population, automation seems imperative to satisfy the growing need for crop production \citep{fao2018the}.  Precision agriculture offers both financial and environmental benefits, leading toward a sustainable model. The literature on agricultural remote sensing applications addresses a variety of challenges such as biomass and yield estimation, stress and development monitoring, vegetation mapping, and crop classification \citep{Weiss2020, Khanal2020, Cisternas2020}.
\subsection{Study Domain}
\label{s:study-domain}
This study belongs to the field of crop phenological development monitoring, in which the continuous development process is partitioned into a sequence of discrete phenological stages. In the case of annual crops, these stages typically encompass planting, emergence, leaf initiation, floral initiation, flowering, and maturity. The specific documentation of these stages may vary slightly depending on the crop species and the protocol employed in the study. While stage transitions possess clear biological demarcations, seed planting (initializing stage) and crop harvesting (terminating stage) are established based on agronomic criteria \citep{sadras2021crop}. It is essential to distinguish crop development from crop growth; the former concerns physiological and morphological changes and, therefore, is measured using ordinal variables, while the latter focuses on quantitative characteristics such as yield \citep{sadras2021crop}. \par
The interest of this study resides in large-scale monitoring. In this context, development is expressed as the percentage of crops that occupy each phenological stage at a particular moment in time, referred to as crop progress. Accurate monitoring of crop progress holds significant importance for informed agricultural decision-making \citep{sadras2021crop}.  However, effectively monitoring crop progress across expansive areas necessitates the meticulous tracking of individual fields, a task that is both costly and time-intensive.
\subsection{State of the Art}
\label{s:state-of-the-art}
Several studies have addressed the topic of crop progress monitoring. A natural approach involves performing  pixel-level inference and then calculating the stage percentages over the area of interest. However, this approach entails individual pixel errors. Furthermore, as the number of pixels increases, so does the computational time required, forcing a balance between model complexity and area extent. Applications in this category employ simple modeling mechanisms and restrict to a small number of stages. The studies of \citet{gao2017toward} and \citet{seo2019improving} constitute such examples, using inflection and threshold methods on smoothed NDVI time series to determine three phenological stages for corn and soybeans, corresponding to the start, peak, and end of the season. \par
Hidden Markov Models (HMMs) have also been employed in the study of crop progress monitoring. \citet{shen2013hidden} constructed a non - homogeneous Gaussian HMM. The predictive features used were thermal time, NDVI, and fractal dimension. These features were combined in a vector, assumed to follow a multivariate normal distribution conditional on a hidden state representing the phenological stage occupied by the crops. The initial distribution probabilities were estimated by averaging the historical data percentages. This methodology was further improved by \citet{ghamghami2020a}, who estimated the initial probabilities by fitting a Gamma distribution with season-dependent hyperparameters to capture the data inter-annual variability. Both methodologies were applied to corn. \par
\subsection{Study Innovation and Application}
\label{s:study-innovation}
A prevailing methodology for ordinal variable modeling that has not been previously applied in crop phenology is the Cumulative Link Model (CLM), which works under the assumption that the crop progress vector follows a Multinomial distribution \citep{peterson1990partial}. Furthermore, this study presents a new type of CLM based on the Multivariate Binomial distribution. Both approaches demonstrate computational efficiency and versatility, enabling real-time prediction of crop progress across various crop types. An important advantage of the Multivariate Binomial CLM is that it can be implemented using any software capable of Binomial regression modeling. The proposed methodology provides a straightforward means to model the crop progress dynamics and estimate the impact of calendar and thermal time, as well as biological parameters that determine stage completion requirements. Furthermore, an alternative approach using the Normalized Difference Vegetation Index (NDVI) is investigated. \par
The study addresses two problems, new season prediction, and completed season fitting. In the first case, the models provide real-time predictions utilizing the calendar and thermal time. In the latter case, random effects are incorporated to account for the inherent inter-annual variability, thereby facilitating accurate fitting and interpolation. In the present application, this approach enables the transformation of weekly progress observations into daily ones. \par 
The case study focuses on the US state of Nebraska, an area of approximately $200,000$km$^2$ with a highly developed agriculture sector. The major crops cultivated in Nebraska are corn, soybeans, sorghum, winter wheat, millet, oats, dry beans, and alfalfa, the collective production value of which surpassed the $16$ billion US dollars in 2021 \citep{nass2023}. The time interval of the study is 20 years long, from 2002 to 2021. The analysis was executed with the R programming language \citep{r2023}. Focusing on research dissemination, an ecosystem of R packages was developed under the name \textit{Ages of Man}. This project can be used to apply the methodology to all available crops for all US states. Technical details concerning the packages and illustrative code examples can be found in the relevant documentation (see Supporting Materials) and will not be further discussed in the paper.
\section{Materials}
\label{s:materials}
This section concerns the materials used in the study, offering details on data acquisition and processing. It is divided into three subsections, covering agricultural, environmental, and satellite data. All data used are open-source and can be acquired directly from their respective provider or via R with the package \texttt{agesofman}. \par
Conciseness of the text invokes the adaptation of simple and coherent notation. The set of consecutive integers up to $N$ will be denoted as $[N]:=\{1, 2, \dots, N\}$. Three indices will be employed throughout the study: $i \in [I]$ for the growing season, $j \in [J]$ for the time-step within a season, and $k \in [K]$ for the phenological stage. The exponent $n \in [N]$ will denote the n-th plant. Graphs in the main text are only presented for corn; the corresponding graphs for all crops are included in the supplementary information.
\subsection{Crop Progress}
\label{s:phenological-development}
\begin{figure}
\centering
    \centerline{\includegraphics[width=88mm]{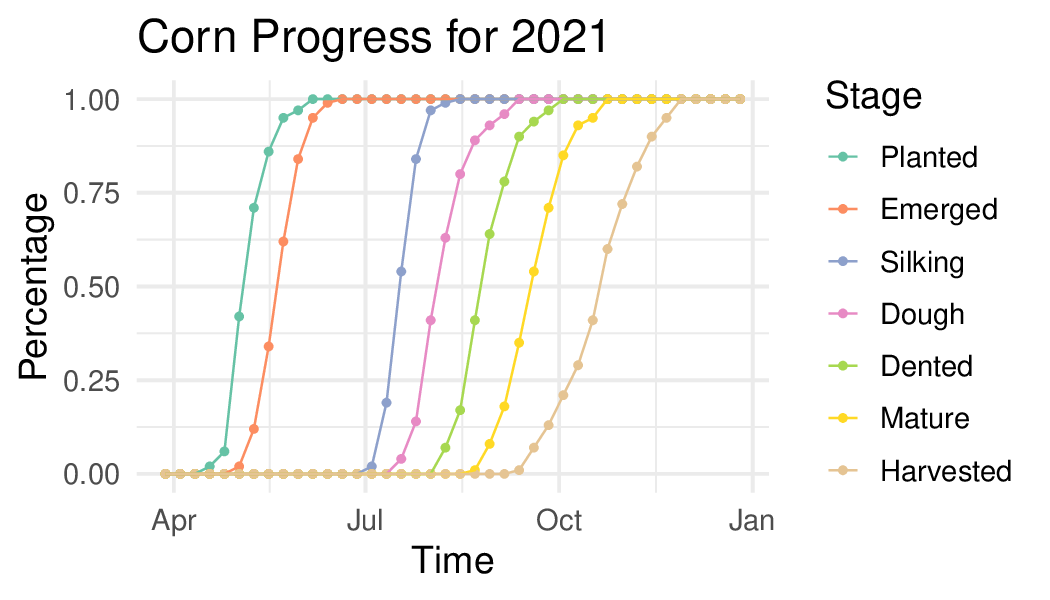}}
    \caption{Corn progress for 2021. The data are indicated with color-coded points for each stage. Lines are used to connect the data points for illustration purposes.}
    \label{fig:corn_progress}
\end{figure}
\begin{definition} \label{def:crop-stage-vector}
The \textit{stage vector} of the $n$-th plant for season $i$ and time-step $j$ is defined as the vector $\bmath{e_{ij}^n}\in \left\{0, 1\right\}^K$ such that $e_{ijk}^n=1$ if the plant has reached stage $k$ at the given moment, and $e_{ijk}^n=0$ otherwise. The \textit{stage} $s^n_{ij} \in [K]$ is defined as $s^n_{ij} = \sum_{k = 1}^K e^n_{ijk}$. The \textit{progress vector} $\bmath{y_{ij}}\in \left[0, 1\right]^K$ is defined as the stage vector average with respect to the plants, that is $\bmath{y_{ij}} = \sum_{n=1}^N \bmath{e_{ij}^n} / N$.
\end{definition}
Crop development is a continuous process, but dividing it into phenological stages allows agricultural practitioners to monitor specific development patterns and therefore optimize their cultivation strategies, simplifying decision-making \citep{sadras2016crop}. The crop progress vector shows the percentage of crops that have reached stage $k$. The nature of crop development forces $y_{ijk}$ to increase with respect to time-step $j$ and decrease with respect to stage $k$. It is straightforward to verify the connection $E(y_{ijk}) = E(e^n_{ijk}) = P(s^n_{ij} > k-1)$. \par
The National Agricultural Statistics Service (NASS) of the United States Department of Agriculture (USDA) conducts crop progress surveys \citep{nass2023}, which provide weekly updates of crop progress through various phenological stages in the growing season (Definition \ref{def:crop-stage-vector}). Figure \ref{fig:corn_progress} illustrates the corn progress data for 2021. NASS also provides a geospatial data product called Cropland Data Layer (CDL). It is a geo-referenced, crop-specific land cover map created annually for the continental United States, using satellite and ground truth data. CDL pixels are nominal categorical variables that take values such as corn, soybeans, or winter wheat. Even though CDL is an agricultural product, it includes other categories such as forest, urban, water, or pasture to be a complete cover map. \par 
This study concerns 8 crops, specifically corn, sorghum, soybeans, winter wheat,  oats, dry beans, alfalfa, and millet. Table \ref{table:crop-summary} summarizes information about the number of pixels and seasons available. All crops considered in the study are annual, meaning that their life cycle is completed within one year, therefore terms \textit{season} and \textit{year} will be used interchangeably throughout the text.
\subsection{Thermal Time}
\label{s:thermal-time}
\begin{figure}
\centering
    \centerline{\includegraphics[width=88mm]{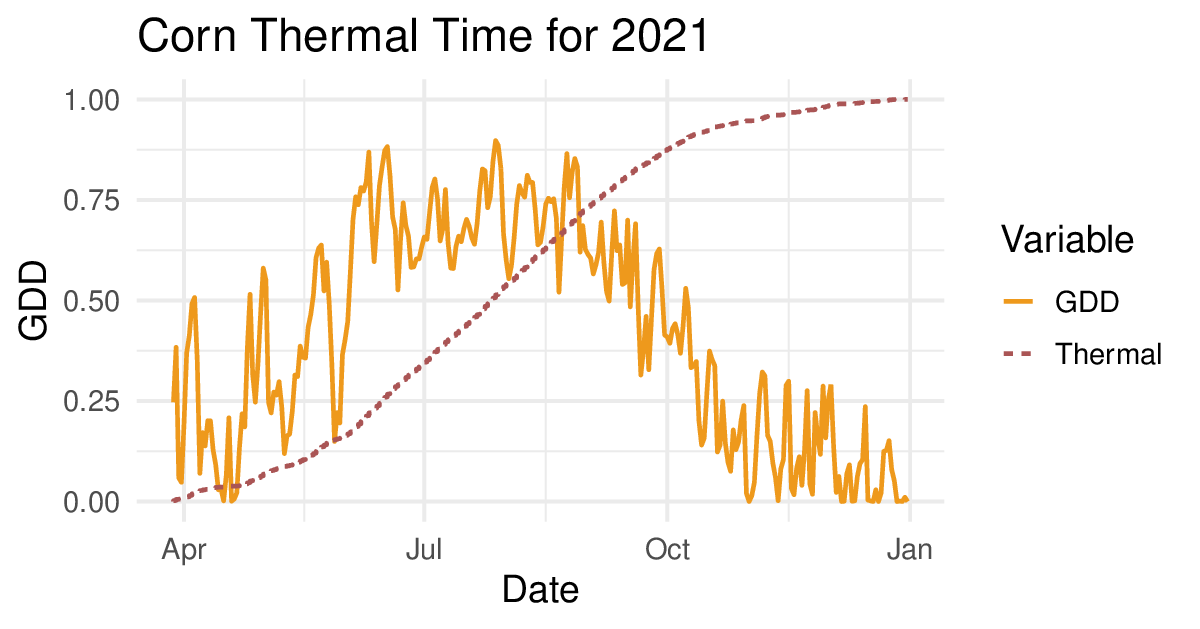}}
    \caption{Daily growing degree days (solid orange line), and thermal time (dashed red line), averaged over corn fields for 2021. Thermal time is scaled to reach a maximum of 1 for illustration purposes.}
    \label{fig:corn-thermal}
\end{figure}
Plants have extraordinary capabilities to haste or slow their development in accordance with environmental conditions. Specifically, the development rate of a crop is considered to be proportional to a type of temperature-weighted time, called thermal time \citep{sadras2016crop}. It is important not to confuse the two variables used in this study. The \textit{calendar time} will be denoted by $t_{ij}$, measured in $\text{d}$ (days), and \textit{thermal time} will be denoted by $\tau_{ij}$, measured in $^\circ C \cdot \text{d}$. Thermal time can be modeled in several ways, revolving around the fact that crops develop within a range of temperatures, outside of which development stops. Key (cardinal) temperatures usually include a base $T_b$, an optimal $T_o$, and a ceiling $T_c$ one. This study follows the heat stress modeling approach \citep{gilmore1958heat, cross1972prediction}, which allows for correcting temperatures exceeding $T_o$ (Definition \ref{def:thermal-time}).
\begin{definition} \label{def:thermal-time}
    Let $T_b$, $T_o$, and $T_c$ be the crop cardinal temperatures. The \textit{truncated average temperature} of season $i$, day $j$ is defined as
    \[ 
     T_{av}(i, j) = \frac{\max\{T_{\min}(i, j), T_b\} + \min\{T_{\max}(i, j), T_c\}}{2}.
    \]
    The corresponding \textit{growing degree day} (GDD) is defined as
    \[
    \text{GDD}(i,j) := cf_{Tr}\left(T_{av}(i, j); T_b, T_o, T_c\right),
    \]
    where $f_{Tr}\left(x; l, m, u\right)$ is the density function of the Triangular distribution with lower limit $l$, mode $m$, and upper limit $u$, and $c:=f^{-1}_{Tr}\left(m; l, m, u\right)$ is a normalization constant so that GDD takes values in $[0, 1]$. The \textit{thermal time} of season $i$ and day $j$ is defined as the accumulated GDD, $\tau_{ij} := \sum_{l=1}^{j} \text{GDD}(i,l)$.
\end{definition}
In this study, temperature data are obtained from Daymet, a research product of the Environmental Sciences Division at Oak Ridge National Laboratory (ORNL). Daymet provides daily, $1\text{km}\times1\text{km}$ gridded weather variable estimates \citep{daymet2023}, including minimum and maximum $2m$ air temperature ($^\circ C$), from which the growing degree days can be inferred. Finally, the CDL can be used as a crop mask to produce a single, average time series for each crop over the whole area of interest (Figure \ref{fig:corn-thermal}). The cardinal temperatures used for the crops under study can be found in Table \ref{table:crop-summary} \citep{ferrante2018agronomic, ong1985response}.
\begin{table}[htbp]
\caption{Crop Summary Information} \label{table:crop-summary}
{\begin{tabular*}{\linewidth}{@{}l@{\extracolsep{\fill}}l@{\extracolsep{\fill}}r@{\extracolsep{\fill}}r
@{\extracolsep{\fill}}r@{\extracolsep{\fill}}r@{\extracolsep{\fill}}r@{\extracolsep{\fill}}c@{}}
\Hline
Crop & Scientific Name & Pixels & Seasons & $T_b$ & $T_o$ & $T_c$ \\
\hline
Corn     & Zea mays           & 200231 & 20 & 8  & 30 & 36 \\
Sorghum  & Sorghum bicolor    & 1148   & 20 & 12 & 30 & 36 \\ 
Soybeans & Glycine max        & 95146  & 20 & 10 & 28 & 34 \\
Wheat    & Triticum aestivum  & 19308  & 19 & 2  & 26 & 32 \\
Oats     & Avena sativa       & 281    & 17 & 2  & 26 & 32 \\
Beans    & Phaseolus vulgaris & 1921   & 7  & 10 & 30 & 36 \\
Alfalfa  & Medicago sativa    & 7974   & 4  & 8  & 26 & 36 \\
Millet   & Pennisetum glaucum & 1148   & 4  & 11 & 33 & 46 \\
\hline
\multicolumn{7}{l}{Cardinal temperatures are given in $^\circ C$.}
\end{tabular*}}
\bigskip
\end{table}
\subsection{Normalized Difference Vegetation Index}
\label{s:ndvi}
Vegetation Indices (VIs) are a group of red and infrared radiance functions designed to estimate the chlorophyll levels of vegetation, with Normalized Difference Vegetation Index (Definition \ref{def:ndvi}) being the most common \citep{rouse1974monitoring}. In this study, the new concept of NDVI-based greenup is introduced by replicating the formulation of thermal time. 
\begin{definition} \label{def:ndvi}
Let $\rho_{\text{RED}}\in(0,1)$ and $\rho_{\text{NIR}}\in(0,1)$ represent the surface reflectance averaged over ranges of wavelengths in the visible red ($620-670$ nm) and near-infrared ($841-876$ nm) regions of the spectrum, respectively. The Normalized Difference Vegetation Index (NDVI) of season $i$ and day $j$ is defined as:
\[
    v_{ij} := \frac{\rho_{\text{NIR}}(i, j) - \rho_{\text{RED}}(i, j)}{\rho_{\text{NIR}}(i, j) + \rho_{\text{RED}}(i, j)},
\]
where $\rho_{\text{RED}}, \rho_{\text{NIR}} \in (0,1)$, resulting in $v_{ij} \in (-1,1)$. The \textit{greenup} of season $i$ and day $j$ is defined as the accumulated NDVI, $g_{ij} := \sum_{l=1}^{j} v_{il}$.
\end{definition}
High reflectance in the near-infrared and low in the red band results in high NDVI values. This combination is typical of vegetation, while non-vegetated areas, including bare soil, water, and most construction materials, acquire much lower NDVI values. \par
\begin{figure}
\centering
    \centerline{\includegraphics[width=88mm]{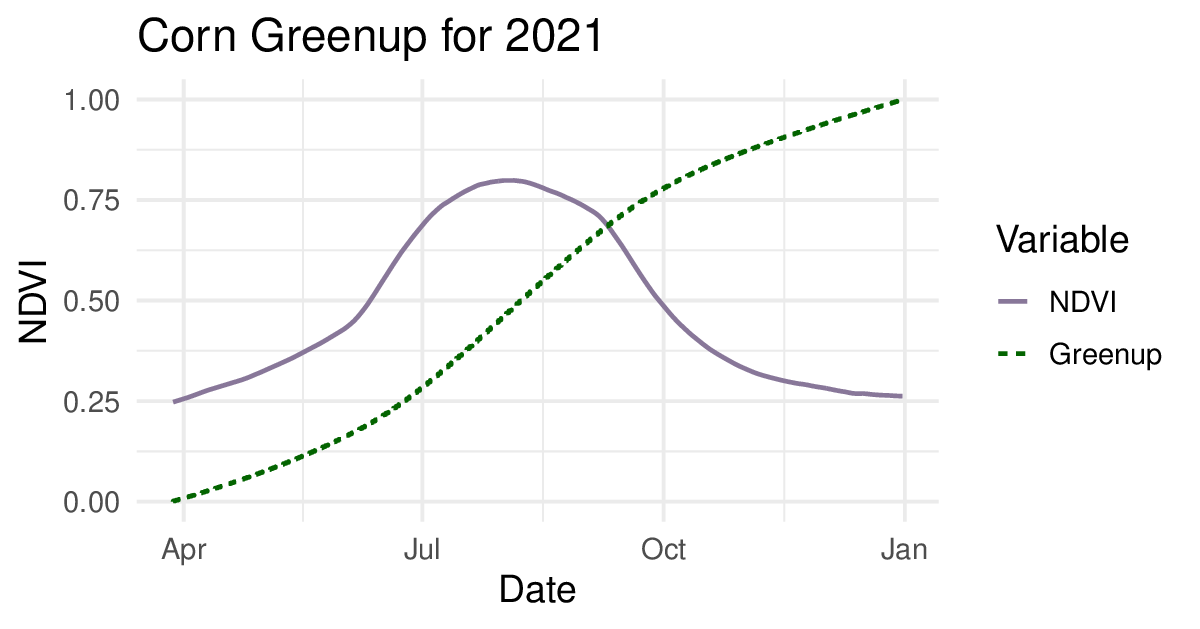}}
    \caption{Daily smoothed NDVI (solid purple line) and greenup (dashed green line), averaged over corn fields for 2021. Greenup is scaled to reach a maximum of 1 for illustration purposes.}
    \label{fig:corn-greenup}
\end{figure}

In this study, NDVI is obtained from the Moderate Resolution Imaging Spectroradiometer (MODIS), a sensor aboard the Terra and Aqua satellites \citep{modis2002moderate}. The MOD09GA Version 6 product provides daily $500\text{m}\times500\text{m}$ gridded estimates of the surface spectral reflectance corrected for atmospheric conditions such as gasses, aerosols, and Rayleigh scattering. Satellite data are prone to noise mainly caused by clouds, which obscure vision and hide the areas of interest.  This product is meant to be used with the cloud correction masks and other important quality and viewing geometry information, which are also included in MOD09GA \citep{vermote2015modis}. Smoothing techniques can be applied to each pixel, allowing the interpolation of the gaps created by the cloud mask, creating a smooth NDVI time series. In this study, the first-order Whittaker smoother is used \citep{whittaker1922new, eilers2003perfect, geng2014comparison}. Finally, the CDL can be used as a crop mask to produce a single, average time series for each crop over the whole area of interest (Figure \ref{fig:corn-greenup}). \par
It is interesting to witness the progressive nature of the model in terms of the predictive features $t_{ij}$, $\tau_{ij}$ and $v_{ij}$. When only the calendar time is considered, the model establishes a fundamental baseline for crop progress based on the day of the year, referred to as the \textit{Calendar} setting. By including thermal time, the model incorporates the primary driver of crop development, referred to as the \textit{Thermal} setting. The \textit{Greenup} setting can be constructed by incorporating calendar time and greenup, imitating the Thermal setting. Finally, the \textit{Combined} setting includes the calendar and thermal time, as well as the NDVI. All predictive features were standardized (sample mean equal to 0, sample standard deviation equal to 1) before the fitting process to avoid numerical issues. \par
\section{Methods}
\label{s:methods}
This section focuses on elucidating the methodology employed in the study. It begins by establishing fundamental definitions and providing a comprehensive explanation of the Cumulative Linear Model (CLM), along with formulation arguments grounded in explicit biological hypotheses. Subsequently, the section proceeds to outline the methods used for parameter estimation and model evaluation. \par 
The predictor and parameter vectors are denoted by $\bmath{x_{ij}}\in \mathcal{X}\subseteq\mathbb{R}^{m}$ and $\bmath{\theta} \in \bmath{\Theta} \subseteq \mathbb{R}^q$, occasionally broken down to components $\bmath{x_{ij}} = \left(\bmath{w_{ij}}, \bmath{z_{ij}} \right)$ and $\bmath{\theta} = \left(\bmath{\alpha}, \bmath{\beta}\right)$. The conditional expectation $E_{\bmath{\theta}}\left(\bmath{y_{ij}}\,|\,\bmath{x_{ij}}\right)$ will be denoted as $\bmath{m}(\bmath{x_{ij}}, \bmath{\theta})$, or $\bmath{m_{ij}}$ for simplicity. In the text, a common abuse of notation involves the element-wise application of $\log$ and $\text{logit}$ on vector $\bmath{x}$, which is denoted as $\log\bmath{x}$ and $\text{logit}\,\bmath{x}$, respectively. Finally, the term density refers to the Radon-Nikodym derivative, covering both absolutely continuous and discrete distributions.
\subsection{Essential Definitions}
\label{s:essential-definitions}
This study develops parametric models based on linear exponential families, which can be found in any standard statistics textbook such as \citet{wooldridge2010econometric}.  Definitions \ref{def:bcm} and \ref{def:mb} concern the Multinomial and Multivariate Binomial distribution families, respectively.
\begin{definition} \label{def:bcm}
    A conditional linear exponential family is called \textit{Multinomial} with parameters $N\in\mathbb{N}$ (known) and $\bmath{m_{ij}}\in [0,1]^K: \sum_{k=1}^{K} m_{ijk} = 1$, denoted by $\mathcal{M}\left(N, \bmath{m_{ij}}\right)$, if its density takes the form:
    \[
        f\left(\bmath{y_{ij}}|\bmath{m_{ij}}\right) = \binom{N}{N\bmath{y_{ij}}} \exp\left\{N \bmath{y^\top_{ij}} \log \bmath{m_{ij}}\right\},
    \]
    for all $\bmath{\theta} \in \bmath{\Theta}$ and $\bmath{y_{ij}}\in \{0, 1/N, \dots, 1\}^K: \sum_{k=1}^{K} y_{ijk} = 1$.
\end{definition}
By reparameterizing the Multinomial category expectations $m_{ijk}$ into their remaining sums $m^\star_{ijk} := \sum_{l=k}^K m_{ijl}$, the Backward Cumulative Multinomial distribution $\mathcal{BCM}\left(N, \bmath{m^\star}\right)$ arises. In this case, the density is expressed with expectation differences ($m_{ijk} = m^\star_{ijk} - m^\star_{ijk+1}$). Under this law, $y_{ijk}$ represents the proportion of successes distributed among categories $\geq k$ in $N$ trials, which is exactly what crop progress constitutes. The parameter $m^\star_{ijk}$ holds the expected proportion of successes; note that $m^\star_{ij1} = 1$ a.s. for every $i\in [I]$ and $j\in [J]$. \par
Ordinal variable GLMs can take various forms, contingent upon the specific focus of each application \citep{coull2000random}. For a comprehensive reference encompassing related models, the work of \citet{agresti2010analysis} serves as an invaluable and informative textbook. In the context of crop progress analysis, where the primary interest lies in cumulative percentages, the Cumulative Link Model \citep{peterson1990partial} emerges as a natural choice. A Mixed-Effects form is presented in Definition \ref{def:clmm}, and an explanatory discussion follows.
\begin{definition} \label{def:clmm}
    A parametric expectation model is called \textit{Cumulative Link Model} (CLM) if the conditional expectation of $\bmath{y_{ij}}| \bmath{x_{ij}}$ is such that
    \[
        m^\star_{k}(\bmath{x_{ij}}, \bmath{\theta}) = F\left(\bmath{w^\top_{ij}}\bmath{\alpha_{k}} + \bmath{z^\top_{ij}}\bmath{\beta} \right), \quad k \in 2, \dots, K,
    \]
    where $F: \mathbb{R} \rightarrow [0, 1]$ is the inverse link function. The model will be called \textit{Cumulative Link Mixed-Effects Model} (CLMM) if the conditional expectation of $\bmath{y_{ij}}|\left(\bmath{x_{ij}}, \textbf{a}_{\bmath{i}}, \bmath{b_{k}}\right)$ is such that
    \[
        m^\star_{k}(\bmath{x_{ij}}, \bmath{\theta}, \textbf{a}_{\bmath{i}}, \bmath{b_{k}}) = F\left(\alpha_{k} + \text{a}_{ik} + \bmath{x^\top_{ij}}\left(\bmath{\beta} + \bmath{b}_{k} \right) \right),
    \]
    for $k \in 2, \dots, K,$ where $F: \mathbb{R} \rightarrow [0, 1]$ is the inverse link function and $\textbf{a}_{\bmath{i}} \sim \mathcal{N}\left(\bmath{0}, \bmath{\Sigma_{\textbf{a}}}\right), \bmath{b_{k}} \sim \mathcal{N}\left(\bmath{0}, \bmath{\Sigma_b}\right)$ are the random effects.
\end{definition}
The inverse link function $F$ can be any continuous distribution function. Common choices include the Logistic, Normal, and Cauchy distributions; the respective links $F^{-1}$ are called logit, probit, and cauchit. The first term includes the \textit{nominal} parameters $\bmath{a_k}$ which allow for a different effect of covariates $\bmath{w_{ij}}$ on each category. This is in fact a modeling mechanism that imposes stochastical ordinality; assuming a single effect $a_k$, the ordering $-\infty = a_1 < a_2 < \dots < a_K$ results in $1 = m^\star_{1} > m^\star_{2} > \dots > m^\star_{K}$. The second term includes the \textit{ordinal} parameters $\bmath{\beta}$ that force all categories to share the same impact from the covariates $\bmath{x_{ij}}$. \par
In contrast with the fixed-effects CLM that breaks down the predictors into ordinal and nominal, the mixed-effects version can model all predictors with a fixed ordinal effect ($\bmath{\beta}$) and a random nominal one ($\bmath{b_k}$). Furthermore, seasonal random effects ($\textbf{a}_{\bmath{i}}$) can be added to the threshold parameters ($a_k$), modeling the inter-annual variability. This formulation is indeed more natural and can be used to provide valuable insights into the biological parameters of the model. However, the seasonal random effects require observations of each season and, therefore, cannot be utilized for new season prediction. \par
The CLMM of Definition \ref{def:clmm} only specifies the expectation structure and, therefore, need not be developed under the Multinomial law. This study introduces a CLMM based on the Multivariate Binomial distribution (Definition \ref{def:mb}).
\begin{definition} \label{def:mb}
    A conditional linear exponential family is called \textit{Multivariate Binomial} with parameters $N\in\mathbb{N}$ (known) and $\bmath{m^\star_{ij}}\in [0,1]^K$, denoted by $\mathcal{MB}\left(N, \bmath{m^\star_{ij}}\right)$, if its density takes the form:
    \[
        f\left(\bmath{y_{ij}}|\bmath{m^\star_{ij}}\right) = \exp\left\{N\bmath{y^\top_{ij}} \,\text{logit} \,\bmath{m^\star_{ij}}\right\} \prod_{i=1}^K\binom{N}{Ny_{ijk}}(1-m^\star_{ijk})^N,
    \]
    for all $\bmath{\theta} \in \bmath{\Theta}$ and $\bmath{y_{ij}}\in \{0, 1/N, \dots, 1\}^K$.
\end{definition}
Even though the two density models share the same expectation structure, the interpretation of $\bmath{y_{ij}}$ differs between the two. In the BCM-CLM, $\bmath{y_{ij}}$ is \textit{deterministically} ordered with respect to the category index $k$, in contrast to the MB-CLM, which assumes conditional independence of the categories. A concise explanation of the circumstances under which each of the two distributions is most suitable is provided at the conclusion of subsection \ref{s:model-formulation}. Irrespectively of the distribution chosen, the estimators retain their asymptotic properties provided the CLMM is correctly specified (model robustness discussion following Theorem \ref{th:mple}). It should be noted that the MB-CLM has an implementation advantage due to its likelihood form, allowing any software programmed to perform classic Binomial regression to accommodate the MB-CLM model as well.
\subsection{Model Formulation}
\label{s:model-formulation}
Cumulative Link Models can be formulated by employing standard latent variable arguments, which offer a comprehensive framework for the ordinal and nominal effects \citep{wooldridge2010econometric}. These principles can be seamlessly applied to the realm of crop development, thereby facilitating a systematic understanding of the phenomenon. To this end, two key assumptions are introduced. First, Assumption \ref{a:dev-stage} formalizes the inherent connection between a crop's development and its corresponding phenological stage. Second, Assumption \ref{a:dev-pred} addresses the relative nature of crop development and its correlation to thermal time. Both assumptions utilize a latent variable $d^n_{ij}\in\mathcal{D}\subset \mathbb{R}$, which represents the remaining development of the $n$-th plant for season $i$ and time-step $j$ until its harvest. \par 
\begin{assumption} \label{a:dev-stage}
    The remaining development $d^n_{ij}$ of a plant and its phenological stage $s^n_{ij}$ are connected by the threshold parameters $a_k\in A \subseteq \mathbb{R}$, such that
    \[
        s^n_{ij} \geq k \Longleftrightarrow d^n_{ij} \leq \alpha_{k}, \quad k\in 2, \dots, K.
    \]
\end{assumption}
Thresholds $a_k$ regulate the transmission from one stage to the next one and are strictly increasing with respect to $k$. The difference $\delta_{k}:=\alpha_{k} - \alpha_{k-1}$ is called the \textit{requirement} to transit from stage $k-1$ to $k$ and is of particular interest. Assumption \ref{a:dev-stage} can be relaxed using variable thresholds. Replacing $\alpha_{k}$ by $\bmath{w^\top_{ij}}\bmath{\alpha_{k}}$ allows the predictors $\bmath{w_{ij}}$ to affect the thresholds. In the context of crop development, this could be used to allow thermal time to have a different effect on each stage.
\begin{assumption} \label{a:dev-pred}
    The remaining development $d^n_{ij}$ of a plant and the corresponding environmental factors $\bmath{z_{ij}}$ are connected by the parameters $\bmath{\beta}\in\bmath{B} \subseteq \mathbb{R}^{q}$, such that
    \[
       d^n_{ij} = -\bmath{z^\top_{ij}}\bmath{\beta} + \epsilon^n_{ij},
    \]
    where $\epsilon^n_{ij}$ are the random errors following a distribution with cumulative distribution function $F$, such that $E\left(\epsilon^n_{ij}\right) = 0$ and $Var\left(\epsilon^n_{ij}\right) = \sigma^2$.
\end{assumption}
Assumption \ref{a:dev-pred} highlights the multifaceted nature of the developmental process and its association with thermal time. The negative sign is used to imply that, under a positive $\beta$, increasing $z_{ij}$ reduces the remaining development $d_{ij}$. Under Assumptions \ref{a:dev-stage}, \ref{a:dev-pred}, and the discussion after Definition \ref{def:crop-stage-vector}, it is straightforward to connect the crop progress vector $\bmath{y_{ij}}$ to the predictor vector $\bmath{x_{ij}}$, completing the formulation of the Cumulative Link Model:
\begin{align*}
    E_{\bmath{\theta}}(y_{ijk}\, | \, \bmath{x_{ij}}) &= P_{\bmath{\theta}}\left(s^n_{ij} \geq k\, | \, \bmath{x_{ij}}\right) = P_{\bmath{\theta}}\left(d^n_{ij} \leq \bmath{w^\top_{ij}}\bmath{\alpha_{k}}\, | \, \bmath{x_{ij}}\right) \\
    &= P_{\bmath{\theta}}\left(-\bmath{z^\top_{ij}}\bmath{\beta} + \epsilon^n_{ij} \leq \bmath{w^\top_{ij}}\bmath{\alpha_{k}}\, | \, \bmath{x_{ij}}\right) \\
    &= P_{\bmath{\theta}}\left(\epsilon^n_{ij} \leq \bmath{w^\top_{ij}}\bmath{\alpha_{k}} + \bmath{z^\top_{ij}}\bmath{\beta}\, | \, \bmath{x_{ij}}\right) \\
    &= F\left(\bmath{w^\top_{ij}}\bmath{\alpha_{k}} + \bmath{z^\top_{ij}}\bmath{\beta}\right).
\end{align*} 
The present application exemplifies a classic panel data framework comprising a collection of independent time series. Specifically, inter-season observations can be regarded as independent due to the annual nature of the crops under investigation. In contrast, intra-season observations cannot be treated as independent since they originate from the same crop population. In the context of completed season fitting, incorporating random effects enables the model to capture the underlying patterns and variations within and between seasons. Therefore, a single random intercept is added for each season and stage, $\textbf{a}_{\bmath{i}} \sim \mathcal{N}\left(\bmath{0}, \bmath{\Sigma_{\textbf{a}}}\right)$. Furthermore, the effect of calendar and thermal time can have random fluctuations for each stage, $\bmath{b_{k}} \sim \mathcal{N}\left(\bmath{0}, \bmath{\Sigma_b}\right)$, resulting in the CLMM of Definition \ref{def:clmm}. \par
As discussed in Section \ref{s:study-domain}, the specific stages documented are contingent upon the study protocol. This feature is of particular importance from the aspect of mathematical modeling. Biologically defined stages, such as emergence or maturity, exhibit \textit{deterministic} stage transitions, naturally fitting to the BCM-CLM. However, stages can also be established from an agronomic perspective. Specific agricultural practices like pruning or the application of fertilizers, while \textit{stochastically} sequenced within the biological cycle, need not adhere to deterministic progression, a characteristic of the Multivariate Binomial distribution. Within the context of this study, the Harvested stage falls into this category, as crops destined for livestock consumption may not necessarily require attainment of maturity prior to harvesting, which stands in contrast to those intended for human consumption \citep{sadras2016crop}.
\subsection{Parameter Estimation}
\label{s:parameter-estimation}
Specifying the full conditional density of $\bmath{y_i}|\bmath{x_i}$ would be rather complicated, whereas the partial density of $\bmath{y_{ij}}|\bmath{x_{ij}}$ is straightforward to formulate. Therefore, the inference is based on partial likelihood, introduced in Definition \ref{def:mple} \citep{cox1975partial}. Theorem \ref{th:mple} summarizes the asymptotic properties of maximum partial likelihood estimators, a proof of which can be found in \citet{newey1994large}. An excellent textbook covering the methodology used in panel data settings is \citet{wooldridge2010econometric}. \par
\begin{definition} \label{def:mple}
 Let $\left\{\left(\bmath{y_{i}}, \bmath{x_{i}}\right), i \in [I]\right\}$, be a random sample with observations $\bmath{y_{ij}}$, $\bmath{x_{ij}}, j \in [J]$. Assume that there is a correctly specified parametric model for the densities $f\left(\bmath{y_{ij}} \mid \bmath{x_{ij}};\bmath{\theta}\right)$. The partial (or pooled) log-likelihood is defined as
\[
    \ell(\bmath{\theta}) = \sum_{i=1}^I\sum_{j=1}^J\ell_{ij}(\bmath{\theta}) = \sum_{i=1}^I\sum_{j=1}^J \log f\left(\bmath{y_{ij}} \mid \bmath{x_{ij}};\bmath{\theta}\right).
\]
The \textit{maximum partial likelihood estimator} (MPLE) is defined as $\widehat{\bmath{\theta}} = \arg\max_{\bmath{\theta}\in\bmath{\Theta}} \ell(\bmath{\theta})$.
\end{definition}
The MPLE can be found using the Fisher score function $\bmath{s}(\bmath{\theta}):=\nabla_{\bmath{\theta}} \ell(\bmath{\theta})$. The score equations $\bmath{s}(\bmath{\theta}) = \bmath{0_q}$ can be solved numerically using the Newton-Raphson algorithm, which in the context of GLMs coincides with the Fisher Scoring algorithm \citep{lange2010numerical}.
\begin{method}
    Let $\bmath{s}:\mathbb{R}^q\longrightarrow\mathbb{R}^q$ be a differentiable function, $\bmath{J_{s}}(\bmath{\theta})$ denote its Jacobian matrix and $\bmath{\theta^\star}$ be a root of the equation $\bmath{s}(\bmath{\theta}) = \bmath{0}$. The \textit{Newton-Raphson algorithm} approximates $\bmath{\theta^\star}$ with the recursive sequence $\bmath{\theta_n}$, such that  
    \[
    \bmath{\theta_{n+1}} = \bmath{\theta_n} - \bmath{J_{s}}\left(\bmath{\theta_n}\right)^{-1}\bmath{s}(\bmath{\theta_n}).
    \]
\end{method}
\begin{theorem} \label{th:mple}
Under regularity assumptions \citep{newey1994large}, the partial maximum likelihood estimator is strongly consistent and asymptotically normal, with
\[
    \sqrt{I}\left(\widehat{\bmath{\theta}}-\bmath{\theta}_0\right)\overset{d}{\longrightarrow}\mathcal{N}_m\left(0,\bmath{A}_0^{-1}\bmath{B}_0\bmath{A}_0^{-1}\right),
\]
where 
\begin{align*}
    \bmath{A}_0 &= -E\left( \nabla_{\bmath{\theta}}^2 \ell_i(\bmath{\theta}_0) \right) = \sum_{j=1}^J E\left( \bmath{s_{ij}}(\bmath{\theta}_0)\bmath{s^\top_{ij}}(\bmath{\theta}_0) \right), \\
    \bmath{B}_0 &=E\left( \bmath{s_i}(\bmath{\theta}_0)\bmath{s^\top_i}(\bmath{\theta}_0)\right) = \bmath{A}_0 + \sum_{j\neq k} E\left( \bmath{s_{ij}}(\bmath{\theta}_0)\bmath{s^\top_{ik}}(\bmath{\theta}_0) \right).
\end{align*}
\end{theorem}
Under the regularity assumptions, it holds that $E\left( \bmath{s_{ij}}(\bmath{\theta}_0) \right)=0$ for $j \in [J]$. Hence, if the score functions $\bmath{s_{ij}}(\bmath{\theta}_0)$ are uncorrelated, the second term of $\bmath{B}_0$ vanishes and $\bmath{A}_0=\bmath{B}_0$. This result is called the unconditional information matrix equality (UIME) and reduces the asymptotic variance presented in Theorem \ref{th:mple} to $\bmath{A}_0^{-1}$. Under UIME, partial and conditional maximum likelihood inference coincide. \par
Robustness calls for a thorough examination of the MPLE asymptotic properties under model misspecification. Assuming that the true conditional density is $p(\bmath{y_{ij}}|\bmath{x_{ij}})$, density misspecification means that no $\bmath{\theta_0} \in \bmath{\Theta}$ exists such that $f(\bmath{y_{ij}}|\bmath{x_{ij}} ; \bmath{\theta_0}) = p(\bmath{y_{ij}}|\bmath{x_{ij}})$. In this case, the focus is adjusted to estimate the $\bmath{\theta^\star} \in \bmath{\Theta}$ that minimizes the Kullback-Leibler divergence, and the estimator of Definition \ref{def:mple} is instead called a \textit{maximum partial pseudo-likelihood estimator}. Linear exponential families offer a key advantage in this direction. Specifically, it can be proved that even if the conditional distribution is misspecified, the correct specification of the conditional mean suffices for the maximum partial pseudo-likelihood estimator to retain its asymptotic properties. Interestingly, this result cannot hold outside of linear exponential families \citep{gourieroux1984pseudo}. \par
When unobserved, random effects are included in the model, the log-likelihood cannot be directly maximized. In order to estimate the parameters, the unobserved effects can be integrated out: $\ell(\bmath{\theta}) = \sum_{i=1}^I\log E_{\textbf{a}_{\bmath{i}}, \bmath{b}}\left[f\left(\bmath{y_{i}} \mid \bmath{x_{i}}, \textbf{a}_{\bmath{i}}, \bmath{b};\bmath{\theta}\right)\right]$. To compute the expectation of interest, Markov Chain Monte Carlo methods can be employed such as in the applications of \citet{trevezas2013sequential, baey2016non, baey2018mixed}. Because this approach is computationally intensive, the Laplace approximation can be used instead \citep{lange2010numerical}.
\begin{method}
Let $\mathcal{D}\subset\mathbb{R}^q$ and $f:\mathcal{D}\rightarrow\mathbb{R}$ be a twice continuously differentiable function, $f\in\mathcal{C}^{2}\left(\mathcal{D}\right)$. If there exists a well-separated point of maximum $\bmath{x_0}\in\text{int}\left(\mathcal{D}\right)$, with a negative definite Hessian matrix $H_f(\bmath{x_0})$, then the following integral approximation
\[
    \int_{\mathcal{D}} e^{f(\bmath{x})} d\bmath{x} \approx \left[\frac{(2\pi)^q}{\text{det}\left(-H_f(\bmath{x_0})\right)}\right]^{1/2}e^{f(\bmath{x_0})},
\]
is called the \textit{Laplace approximation}.
\end{method}
\subsection{Model Evaluation}
\label{s:model-evaluation}
In the context of new season prediction, all possible combinations of link function, ordinal and nominal effects will be compared separately for the crop, setting (Calendar, Thermal, Greenup, Combined), and model type (BCM, MB) in order to perform comparisons under different scenarios. Model performance will be evaluated with the out-of-sample Root Mean Square Prediction Error time series, as well as the average error over the whole season, which can be estimated with Monte Carlo (or leave-group-out) cross-validation \citep{stone1974cross, geisser1975the}, described in Definition \ref{def:rmspe}.
\begin{definition} \label{def:rmspe}
    Let $\bmath{y_{\cdot j}}, \bmath{x_{\cdot  j}}$, $j \in [J]$ be random vectors, $\bmath{\theta}$ the parameter vector and $\bmath{\hat{\theta}}$ the parameter estimate, independent of $\bmath{y_{\cdot j}}, \bmath{x_{\cdot j}}$, $j \in [J]$. The \textit{Root Mean Square Prediction Error time series} (RMSPE) $\bmath{r_{j}}$ and the corresponding \textit{average} $R$ are defined as
    \[
        \bmath{r_{j}} = E_{\bmath{\theta}}\left(\left|\left|\bmath{y_{\cdot j}}-\bmath{m(x_{\cdot j}, \bmath{\hat{\theta}})}\right|\right|^2\right)^{1/2}, \quad R = \overline{\bmath{r}}.
    \] 
    Let $\bmath{y_{ij}}, \bmath{x_{ij}}$, $i \in [I]$, $j \in [J]$ be random vectors and $A_n, B_n$, $n \in [N]$ be random two-set partitions of $[I]$. The \textit{Monte-Carlo estimator} of $\bmath{r_{j}}$ is defined as
    \[
        \hat{r}_j = \frac{1}{N}\sum_{n=1}^N \left(\frac{1}{\left|A_n\right|}\sum_{i\in A_n}\left|\left|\bmath{y_{ij}}-\bmath{m(x_{ij}, \bmath{\hat{\theta}_{B_n}})}\right|\right|^2\right)^{1/2},
    \]
    where $\bmath{\hat{\theta}_{B_n}}$ is the parameter estimate derived from $B_n$. The corresponding estimator of the average error is $\hat{R} = \overline{\bmath{\hat{r}}}$.
\end{definition}
In a single run of the Monte Carlo replications, $75\%$ of the seasons are invested in model training and the remaining $25\%$ are used for testing; in cases where these percentages do not result in natural numbers, such as the 19 seasons of Oats, the training set is favored (15 training and 4 testing seasons). This procedure is repeated independently for all possible train-test set partitions, or $N=500$ times if the total number exceeds $N$ (which is the case for the first five crops). \par
In the context of completed season fitting, a single CLMM model will be constructed and evaluated for each crop. The performance of the model can be checked with the within-sample RMSE; no cross-validation procedure is needed here. The statistical significance of each parameter can be calculated using the Wald z-test. Finally, an interpretation of the predictive variable effects and the threshold parameters is of great value.
\section{Results}
\label{s:results}
This section presents the study's findings, which are divided into two subsections. The first one concentrates on predicting crop progress for new seasons for which no observations have been recorded, while the second one addresses the fitting of completed seasons, employing mixed effects for both seasons and stages.
\begin{table}[htbp]
\caption{Average Root Mean Square Prediction Error in $\%$}
\label{table:model-performance}
\begin{tabular*}{\linewidth}{
r@{\extracolsep{\fill}}
c@{\extracolsep{\fill}}c@{\extracolsep{\fill}}c@{\extracolsep{\fill}}c@{\extracolsep{\fill}}
r@{\extracolsep{\fill}} r@{\extracolsep{\fill}}
c@{\extracolsep{\fill}}c@{\extracolsep{\fill}}c@{\extracolsep{\fill}}c@{\extracolsep{\fill}}
r@{\extracolsep{\fill}} r@{\extracolsep{\fill}}}
\Hline
\multirow{2}*{Setting} & \multicolumn{6}{c}{Corn} & \multicolumn{6}{c}{Sorghum} \\[3pt]
       & \multicolumn{4}{c}{Form} & BCM & MB & \multicolumn{4}{c}{Form} & BCM & MB \\
\hline
Calendar     & p & \wci  &      &       & 13.38 & 13.44 
             & p & \wci  &      &       & 12.34 & 12.61 \\
Thermal      & p & \bci  & \wsq &       & 8.79  &  8.82 
             & p & \bci  & \wsq &       & 11.32 & 11.58 \\
Greenup      & p & \bwci &      & \wtr  & 12.96 & 12.97 
             & p & \bci  &      & \wbtr & 12.77 & 13.34 \\
Combined     & p & \wci  & \bsq & \wtr  & 8.79  &  8.81 
             & p & \bci  & \wsq & \btr  & 11.64 & 11.97 \\
\hline
\multirow{2}*{Setting} & \multicolumn{6}{c}{Soybeans} & \multicolumn{6}{c}{Winter Wheat} \\[3pt]
       & \multicolumn{4}{c}{Form} & BCM & MB & \multicolumn{4}{c}{Form} & BCM & MB \\
\hline
Calendar     & p & \wci &      &      & 12.56 & 12.58 
             & l & \wci &      &      &  8.00 &  8.04 \\
Thermal      & p & \wci & \bsq &      & 11.47 & 11.48 
             & l & \wci & \wsq &      &  6.90 &  6.89 \\
Greenup      & p & \wci &      & \wtr & 12.63 & 12.64 
             & l & \wci &      & \btr &  7.66 &  7.70 \\
Combined     & p & \wci & \bsq & \wtr & 11.16 & 11.19 
             & l & \wci & \wsq & \wtr &  6.61 &  6.60 \\
\hline
\multirow{2}*{Setting} & \multicolumn{6}{c}{Oats} & \multicolumn{6}{c}{Dry Beans} \\[3pt]
       & \multicolumn{4}{c}{Form} & BCM & MB & \multicolumn{4}{c}{Form} & BCM & MB \\
\hline
Calendar     & l & \wci  &       &       & 13.46 & 13.74 
             & p & \wci  &       &       & 17.00 & 16.59  \\
Thermal      & l & \bwci & \wbsq &       & 11.45 & 11.89 
             & p & \wci  & \wsq  &       & 15.25 & 14.91 \\
Greenup      & l & \wci  &       & \wtr  & 12.96 & 13.25 
             & p & \bci  &       & \wtr  & 17.40 & 16.92 \\
Combined     & l & \bci  & \bwsq & \wtr  & 10.96 & 11.41 
             & p & \wbci & \wsq  & \bwtr & 15.49 & 14.94 \\
\hline
\multirow{2}*{Setting} & \multicolumn{6}{c}{Alfalfa} & \multicolumn{6}{c}{Millet} \\[3pt]
       & \multicolumn{4}{c}{Form} & BCM & MB & \multicolumn{4}{c}{Form} & BCM & MB \\
\hline
Calendar     & p  & \wci &      &      & 10.68 & 10.69 
             & l  & \wci &      &      &  8.06 &  8.05  \\
Thermal      & p  & \wci & \wsq &      &  9.59 &  9.64 
             & lp & \bci & \bsq &      &  8.16 &  8.14 \\
Greenup      & p  & \wci &      & \wtr &  9.74 &  9.85 
             & p  & \wci &      & \wtr &  8.24 &  8.27 \\
Combined     & p  & \wci & \wsq & \btr &  9.50 &  9.41
             & p  & \wci & \bsq & \btr &  5.70 &  5.75 \\
\hline
\end{tabular*}
\bigskip
\vspace{0.3cm}
Link: probit (p), logit(l). Effect: nominal (white), ordinal (black). Variable: $t_{ij}$ (\wci), $\tau_{ij}$ (\wsq), $v_{ij}$ or $g_{ij}$ (\wtr).
\end{table}
\subsection{New Season Prediction with Fixed Effects}
\label{s:res-nsp}
Table \ref{table:model-performance} presents the selected models for each crop, covariate setting, and model type. The structure of the selected models is concisely denoted using the following coding scheme: A single letter, p for probit, l for logit, and c for cauchit, denotes the link function. Ordinal effects are represented with black shapes, while nominal effects are represented with white ones. A circle is used for calendar time, a square for thermal time, and a triangle for NDVI or greenup (in the combined and the greenup settings, respectively). In case the forms of the two models disagree, the contrasts are presented side by side (e.g. the Corn Greenup BCM has an ordinal calendar time effect, while the MB has a nominal one). To enhance presentation clarity, the errors are rounded to the fourth decimal place and converted to percentages. \par
Several pertinent notes, supplementary to Table \ref{table:model-performance}, are expounded here, while a complementary discussion comparing the three settings is provided in Section \ref{s:dis-res}. Regarding the link function, the logit and probit links exhibited close competition for the top position, yielding almost identical predictions. Conversely, the cauchit link generally led to significantly higher errors, in some cases $5\%$ higher than the selected model. Under the Calendar covariate setting (i.e. when only the calendar time was used as a predictor), the models that resulted in the lowest errors were the nominal ones, allowing for a distinct effect on each stage. However, it is noteworthy that assuming an ordinal calendar time rarely caused the errors to inflate by more than $0.5\%$. In the other three settings, the optimal structured model varied depending on the crop. This variability can be justified by the fact that many alternative structures achieved similar errors consistently for all crops, especially those including at least one nominal effect. A caveat should be mentioned regarding the limited number of available seasons for the crops: The cross-validation error estimation for Dry Beans, Alfalfa, and Millet was computed on a small number of partitions; hence, the results should be interpreted with caution. \par
\begin{figure}
\centering
    \centerline{\includegraphics[width=88mm]{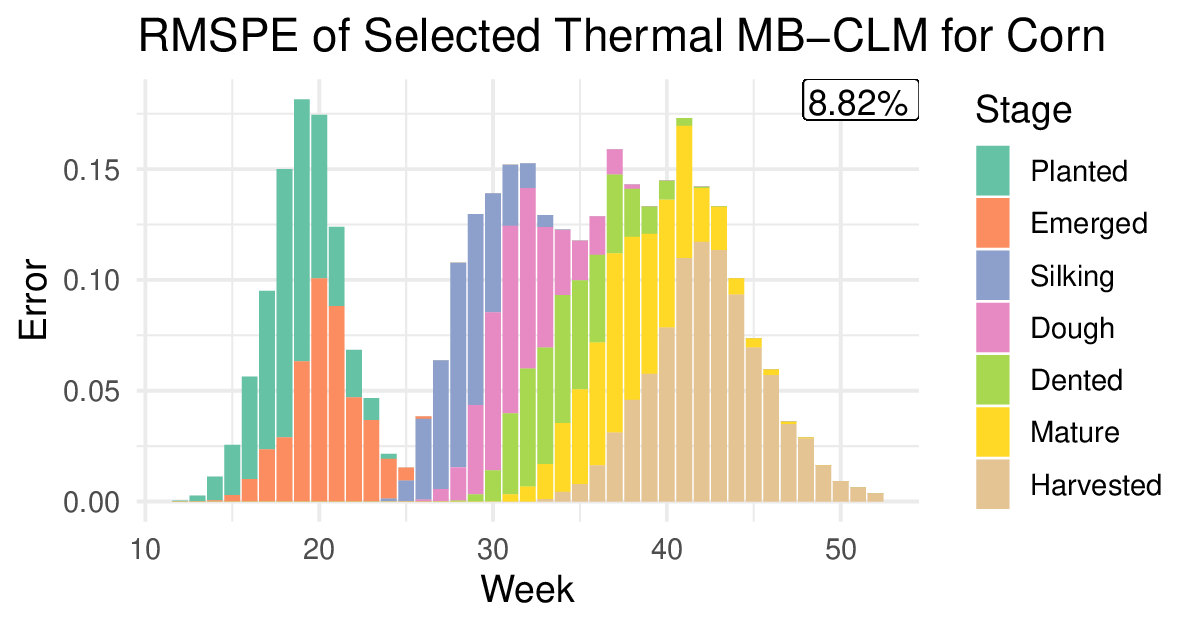}}
    \caption{Root Mean Square Prediction Error, estimated with Monte-Carlo cross-validation on 500 subsets (training 75\% - testing 25\%). The model is the selected MB-CLM for the corn, which facilitates the probit link, calendar time as an ordinal effect, and thermal time as a nominal one. The average error is $8.82\%$. Each stage is denoted with a different color.}
    \label{fig:rmspe_pred}
\end{figure}

The two CLMs yielded very similar results for all crops, links, and covariate settings. Specifically, for the majority of the models, the error difference did not exceed $0.06\%$ (with the exception of Oats and Sorghum). In terms of this application, this gap is particularly small and graphically indistinguishable. Figure \ref{fig:rmspe_pred} presents a stacked bar plot of the $\bmath{\hat{r_{j}}}$ time series for the selected corn MB-CLM under the Thermal setting. The model facilitates the probit link, calendar time as an ordinal effect, and thermal time as a nominal one (denoted by: p \bci\,\wsq). It is of particular interest to break down the average season errors and analyze Figure \ref{fig:rmspe_pred}. This plot should be viewed along with Figure \ref{fig:corn_progress} which shows the crop progress data. It is evident that the uncertainty of each stage is increased during the weeks $y_{ijk}$ attains moderate values around $0.5$, as is expected with categorical variables. Subperiods in which one stage dominates the cultivation result in error deflation, as exemplified by the Emerged stage around Week 25. Finally, it is quite apparent that the planting and harvesting stages generate the highest errors. This pattern holds true across all eight crops, the corresponding graphs of which can be found in the Supporting Materials. \par 
\subsection{Completed Season Fitting with Mixed Effects}
\label{s:res-csf}
\begin{figure}
\centering
    \centerline{\includegraphics[width=88mm]{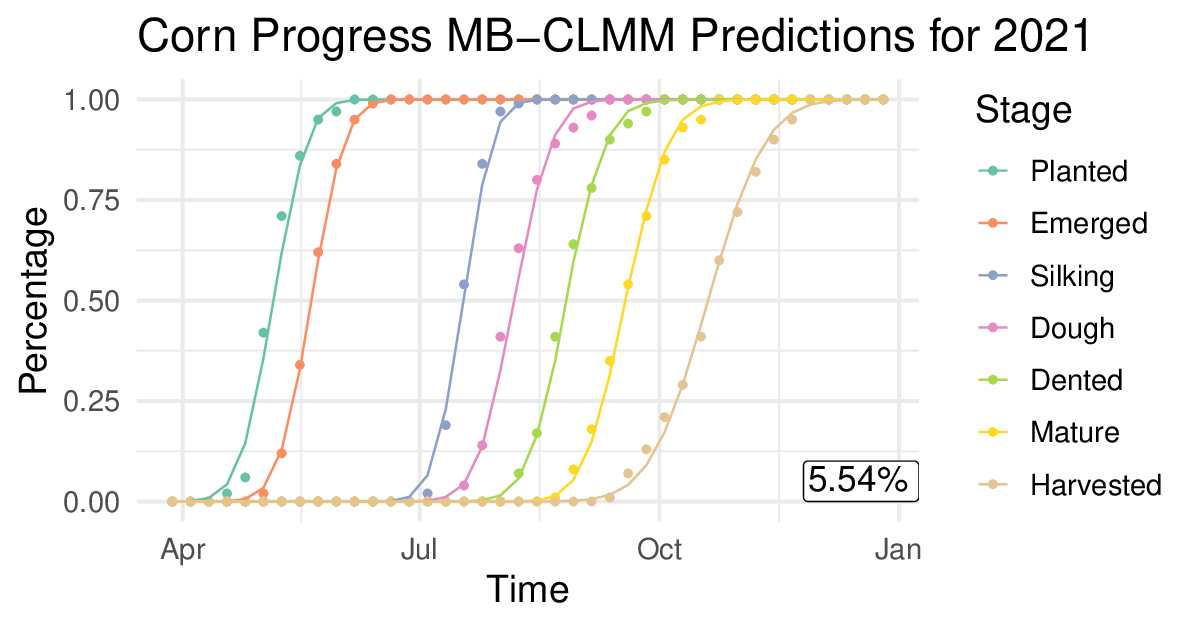}}
    \caption{Corn progress data fitting for 2021. Observations are denoted with points, and fitted values are denoted with a solid line. Each stage is denoted with a different color. The within-sample RMSE of the model averaged over all 20 seasons, is $5.54\%$.}
    \label{fig:corn-mbclmm-pred}
\end{figure}

The results for the corn CLMM fitting are summarized in Table \ref{table:mbclmm-est}. The Table presents the coefficient estimated values, their standard errors, and the corresponding Wald z values, along with their significance level. Furthermore, the last column includes the estimated standard deviations of the random effects, which are clustered by season for the threshold parameters $a_k$, and by stage for the calendar and thermal time. The within-sample RMSE of the model averaged over all 20 seasons, is $5.54\%$. The fitting of the corn progress data for 2021 is presented in Figure \ref{fig:corn-mbclmm-pred}. The corresponding information for all crops can be found in the Supporting Materials. To prevent numerical issues stemming from limited data availability, no random effect for calendar and thermal time was incorporated in the case of Alfalfa and Millet.
\par 
\begin{table}[htbp]
\centering
\caption{Fixed effects Estimates and Random Effect Std.Dev}
\label{table:mbclmm-est}
{\begin{tabular*}{\linewidth}{@{}l@{\extracolsep{\fill}}r@{\extracolsep{\fill}}r
@{\extracolsep{\fill}}r@{\extracolsep{\fill}}r@{\extracolsep{\fill}}r@{\extracolsep{\fill}}c@{}}
\hline
Parameter & Estimate & SE & Wald z & SD \\
\hline
Planted   & 9.728    & 0.070 & 139.756 $\star$  & 0.313 \\
Emerged   & -1.387   & 0.034 & -41.066 $\star$  & 0.146 \\
Silking   & -7.282   & 0.062 & -118.373 $\star$ & 0.275 \\
Dough     & -10.095  & 0.075 & -133.835 $\star$ & 0.339 \\
Dented    & -11.366  & 0.057 & -198.539 $\star$ & 0.256 \\
Mature    & -13.250  & 0.068 & -196.128 $\star$ & 0.305 \\
Harvested & -13.807  & 0.112 & -122.814 $\star$ & 0.509 \\
Calendar   & 5.400    & 0.610 & 8.860 $\star$    & 2.089 \\
Thermal    & 1.357    & 0.352 & 3.855 $\star$    & 1.166 \\
\hline
\multicolumn{6}{l}{Significance level: $\bullet < 0.05, \quad \star < 0.001$}
\end{tabular*}}
\bigskip
\end{table}
\section{Discussion}
\label{s:discussion}
This section serves as a complementary discussion of the presented results, providing additional insights and perspectives. Furthermore, it offers a comprehensive evaluation of the study's scope and findings.
\subsection{In Regards to the Results}
\label{s:dis-res}
Concerning the comparison of the two CLMs presented in this study, it is clear that the results of the two models are similar, both in the best-performing structure (link and ordinality), as well as in the resulting prediction errors. Ultimately, the standard CMB-CLM generally produced slightly lower errors in most of the cases and, therefore, should be used whenever available. However, the MB-CLM has the significant advantage that can be facilitated in any software that performs univariate binomial regression and therefore can be a convenient alternative if the classic CMB-CLM is not available. Ultimately, the purpose of this study is to highlight the benefit of the CLMs as an efficient approach for crop progress modeling. \par
Concerning the settings, it is characteristic that the Calendar models could generally capture the nature of the within-season development, resulting in moderately low errors (Table \ref{table:model-performance}). However, the lack of environmental condition information is clear when compared with their Thermal counterparts. Overall, the best results were presented by the Combined setting which includes the NDVI. However, it is noteworthy that the prediction errors were only slightly decreased with the incorporation of the NDVI and in the case of Sorghum and Dry Beans, it actually caused a slight increase in prediction errors. \par 
The Greenup setting was developed as an alternative approach to the Thermal setting, based on the strong correlation between the two predictors. However, the results clearly demonstrate the superiority of the Thermal setting, as shown in Table \ref{table:model-performance}. Interestingly, incorporating greenup data leads to inflated prediction errors for Sorghum, Soybeans, Beans, and Millet. An explanation for this effect can be found in the GDD-NDVI graphical comparison (Figures \ref{fig:corn-thermal}, \ref{fig:corn-greenup}, and Supporting Materials). The GDD tends to reach 0 when inappropriate cultivation temperatures are observed. In contrast, the NDVI consistently maintains a universal lower bound of around 0.25, which corresponds to non-cultivated soil. As a result, changes in NDVI have a reduced impact on the greenup values due to the lower bound accumulation. Additionally, the smoothing procedure applied to NDVI leads to a loss of fluctuation in the data. A natural hypothesis would be that lower NDVI values caused by high cloud concentration also indicate lower temperatures. Consequently, this NDVI fluctuation could potentially hold significant meaning in the context of monitoring crop phenology. These findings warrant further investigation into alternative smoothing techniques, or even the possibility of not using any smoothing at all, despite it being a standard procedure in NDVI literature.
\subsection{Estimation of the Transition Requirements}
\label{s:dis-requirements}
The threshold parameters $a_k$ serve as breakpoints that delineate different stages. Estimating them acts as a gateway to unlocking the \textit{stage completion requirements} $\delta_k := a_k - a_{k-1}$, calculated in days under specific temperature conditions. These parameters can be utilized in conjunction with the calendar and thermal time effects that indicate the magnitude of the progress achieved by the crops to estimate the time remaining for stage completion. A simple illustrative example is the planting-to-emergence requirement in corn. The two curves are separated by a distance $\delta_3 = a_3 - a_2$ covered gradually by $\beta_1 t + \beta_2 \tau$. Assuming a constant, near-optimal temperature of $25 ^\circ C$ corresponds to a GDD of $0.75$, hence the thermal time can be expressed as $\tau(t) = 0.75 t$. The days required for the transition from planting to emergence is the root $t^\star$ of the equation $\delta_3 = \beta_1 t + 0.75 \beta_2 t$. Vice versa, the transition time could be locked into a desired value and seek the appropriate temperatures to meet the requirement. This example serves to show that the formulation of the CLMM can actively help in decision-making and enhance cultivation planning. The parameters involved in the aforementioned calculations can be either at the population level or, if specific stages and seasons are of concern, at the individual level (i.e. take into account the season random effect for parameters $a$ and the stage random effect for parameters $\beta$). \par
The predictive features utilized in this study are available in real-time on a daily basis. Consequently, the developed methodology can be employed not only to provide crop progress predictions but also to perform interpolation, enriching the weekly time series into daily ones.
\section{Conclusion}
\label{s:conclusion}
In conclusion, this study introduces a crop progress prediction framework based on Cumulative Link Modeling, incorporating calendar, thermal time, and normalized difference vegetation index. The models prove computationally efficient and adaptable to various crops, providing real-time predictions. By utilizing random effects, the CLMM accurately captures inter-annual variability and enables the estimation of the stage completion requirements, as well as the enrichment of weekly progress data into daily estimates. The methodology offers valuable insights into the impact of calendar and thermal time on crop progress, along with essential biological parameters. Notably, the application of CLMs, both fixed and mixed effects, represents a significant contribution to this field. The \textit{Ages of Man} R package ecosystem can be utilized to apply the developed methodology.
\backmatter
\section*{Acknowledgements}
Ioannis Oikonomidis would like to thank the Bodossaki Foundation \citep{bodossaki2023} for funding his doctoral studies.
%
%
\bibliographystyle{biom}
\bibliography{bibliography}
\end{document}


{
\Huge
\centerline{Supporting Materials}
}
\section*{Figure 1: Crop Progress Data}
\begin{figure}[h]
\centering
    \centerline{\includegraphics[width=\textwidth]{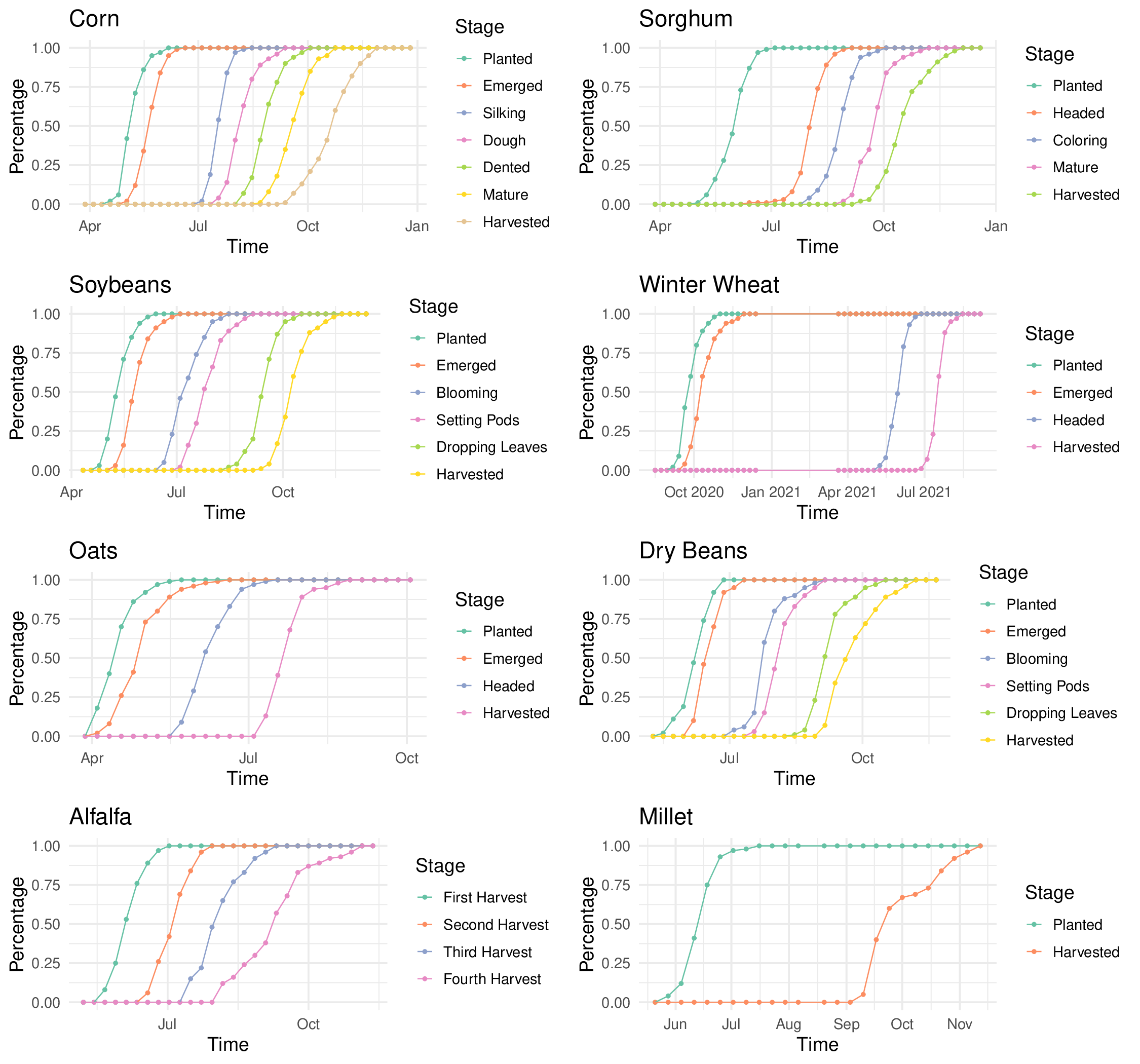}}
    \caption{Crop progress for 2021. The data are indicated with points color-coded for each stage. Dashed lines are used to connect the data points for presentation purposes.}
    \label{fig:crop_progress}
\end{figure}
\newpage
\section*{Figure 2: Crop Thermal Time Data}
\begin{figure}[h]
\centering
    \centerline{\includegraphics[width=\textwidth]{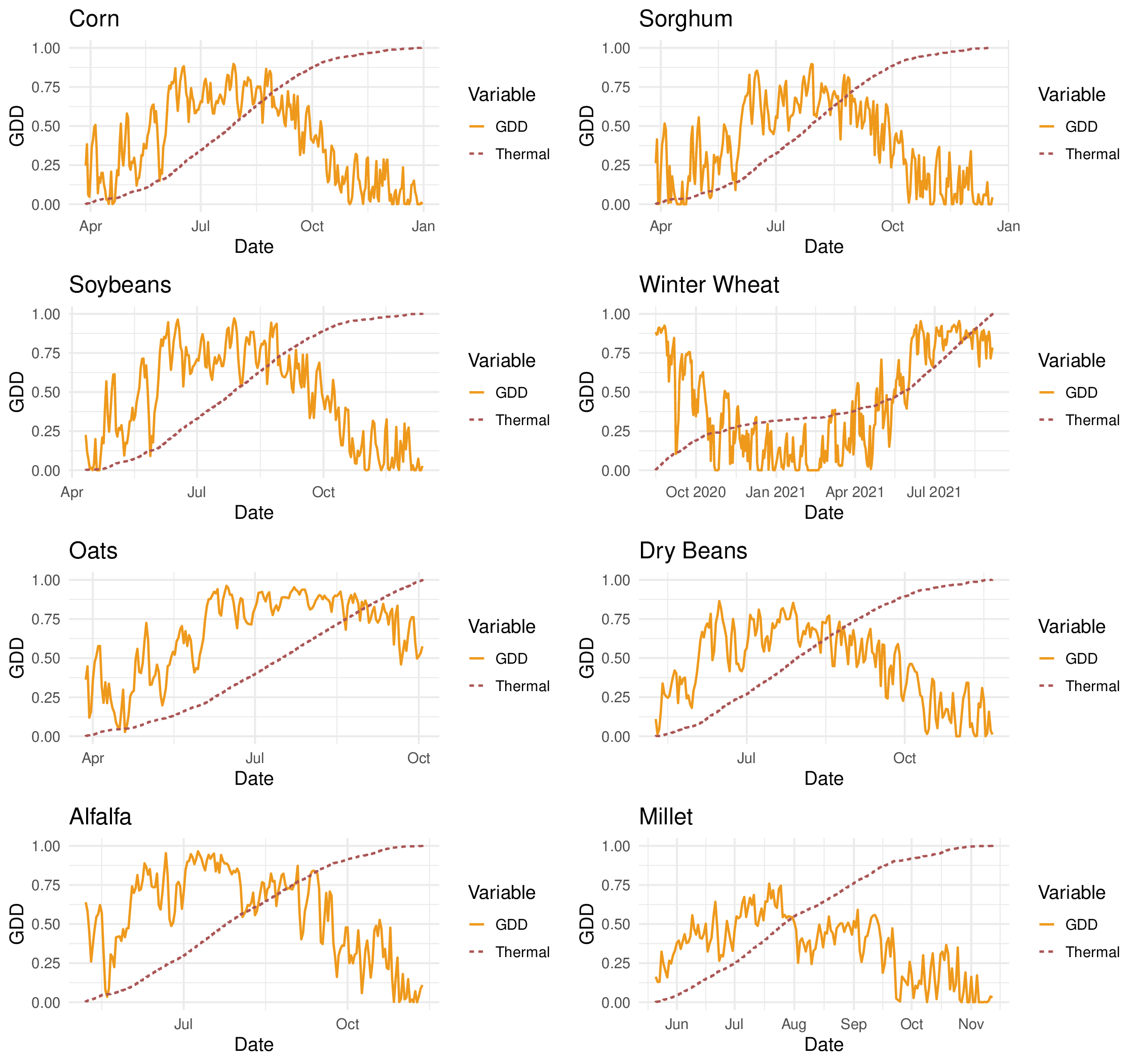}}
    \caption{Daily growing degree days (solid orange line), and thermal time (dashed red line), averaged over crop fields for 2021. Thermal time is scaled to reach a maximum of 1 for illustration purposes.}
    \label{fig:crop_thermal}
\end{figure}
\newpage
\section*{Figure 3: Crop Greenup Data}
\begin{figure}[h]
\centering
    \centerline{\includegraphics[width=\textwidth]{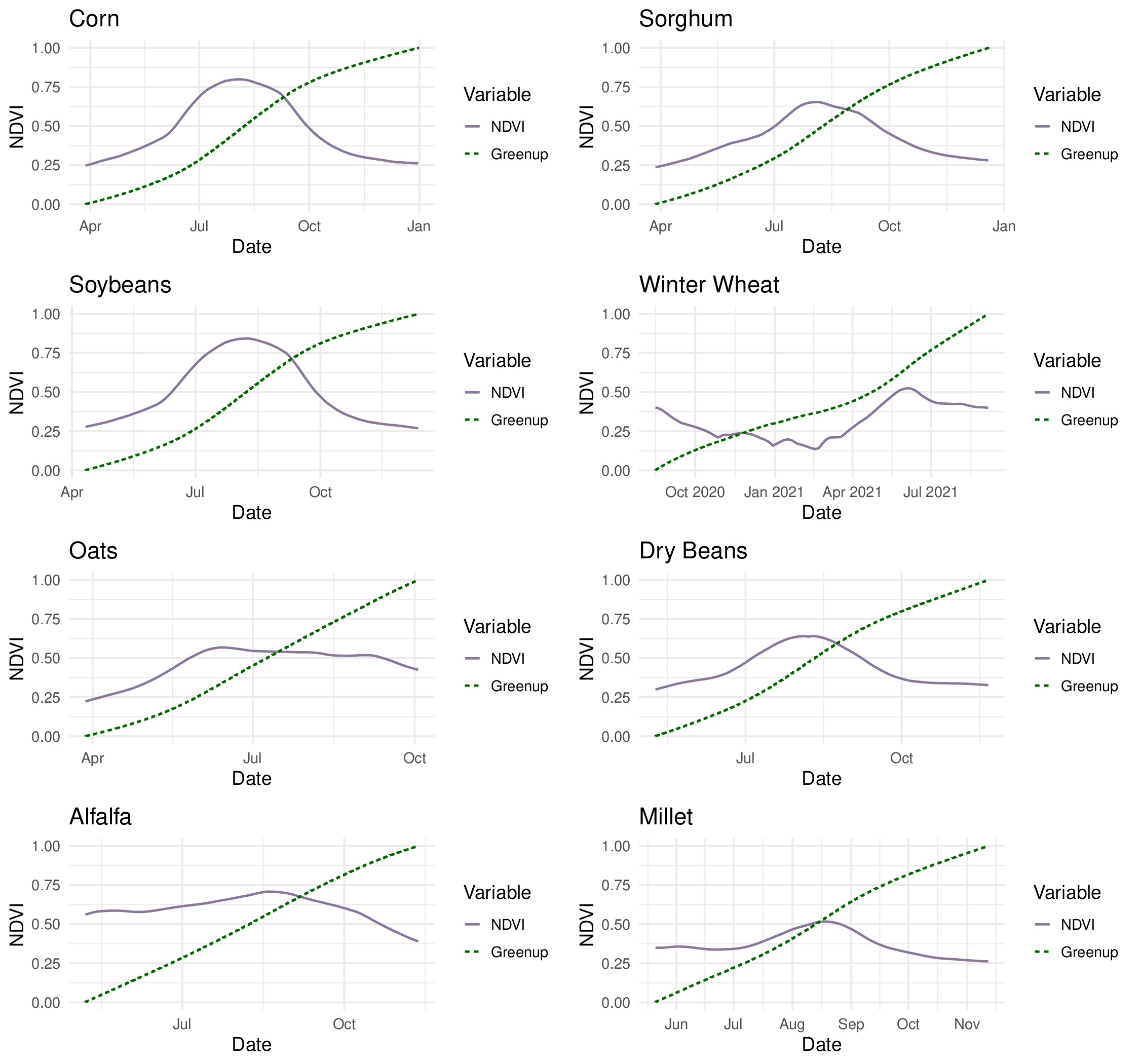}}
    \caption{Daily smoothed NDVI (solid purple line) and greenup (dashed green line), averaged over crop fields for 2021. Greenup is scaled to reach a maximum of 1 for illustration purposes.}
    \label{fig:crop_greenup}
\end{figure}
\newpage
\section*{Figure 2: MB-CLM New Season Prediction Error}
\begin{figure}[h]
    \centering
    \includegraphics[width=\linewidth]{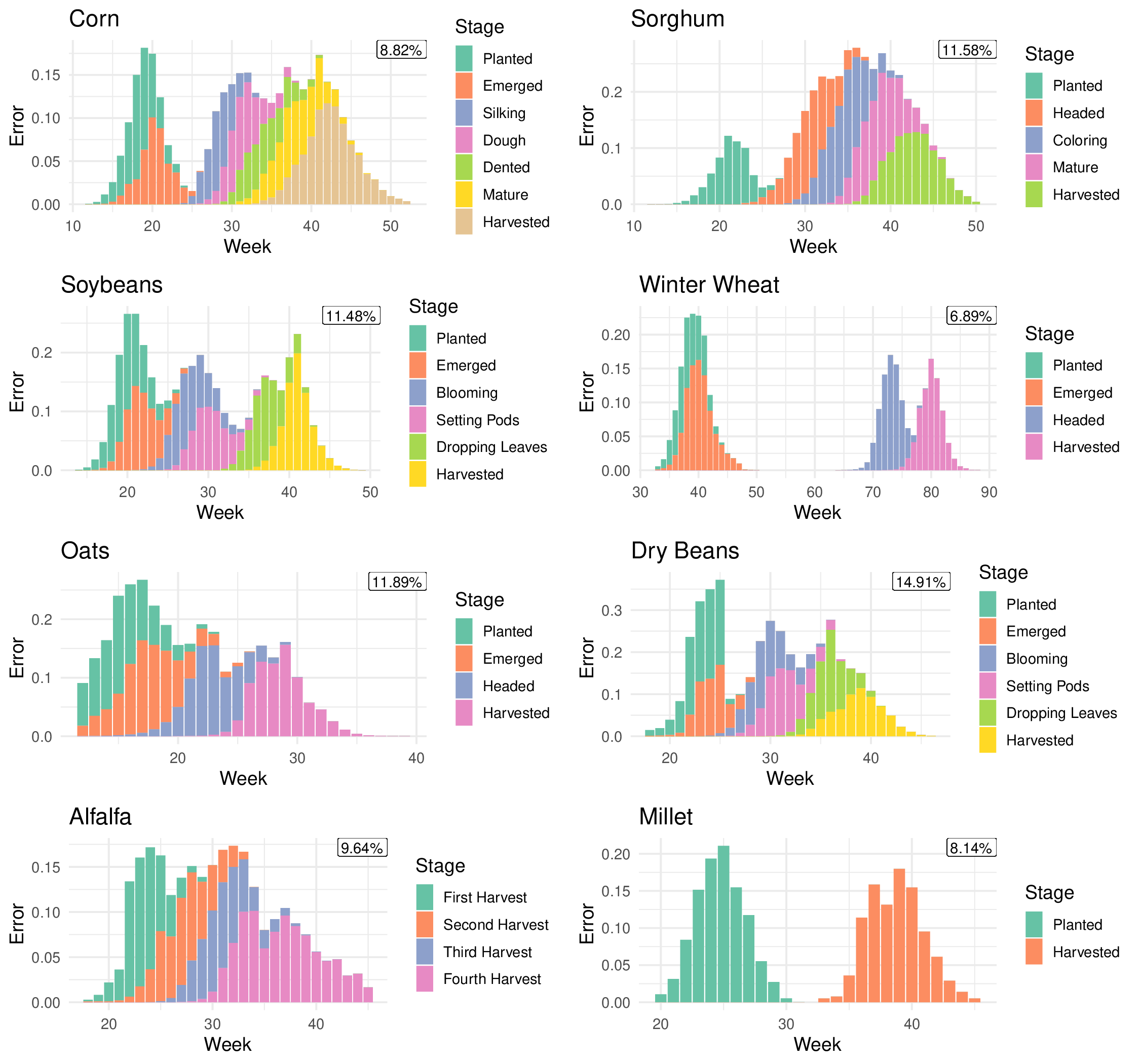}
    \caption{Root Mean Square Prediction Error, estimated with Monte-Carlo cross-validation on 500 subsets (training 75\% - testing 25\%). Each stage is denoted with a different color.}
    \label{fig:rmspe_pred}
\end{figure}
\newpage
\section*{Figure 3: Crop Progress MB-CLMM Fitting for 2021}
\begin{figure}[h!]
\centering
    \centerline{\includegraphics[width=\textwidth]{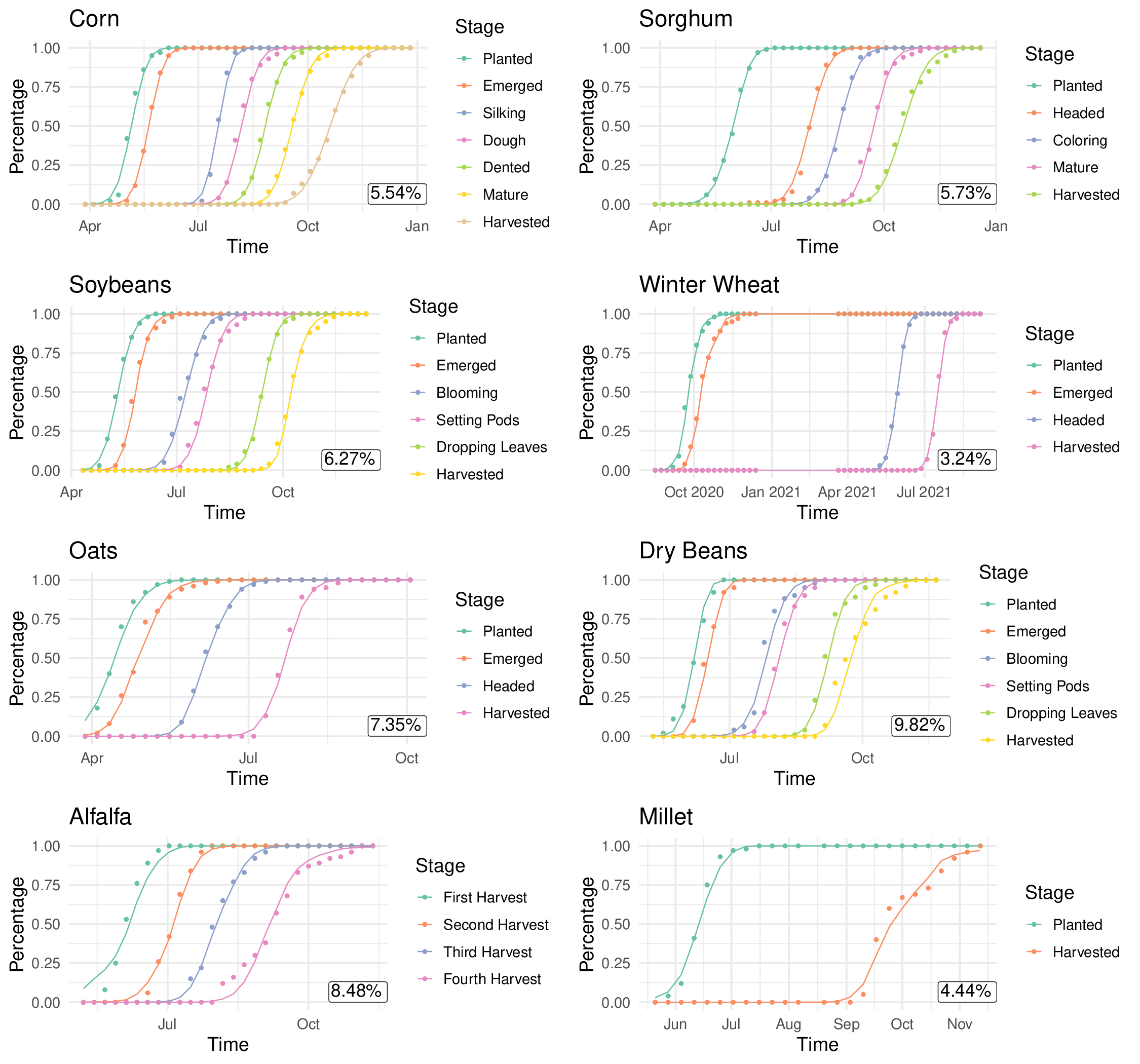}}
    \caption{Crop progress data fitting for 2021. Observations are denoted with points, and fitted values are denoted with a solid line. Each stage is denoted with a different color. The within-sample RMSE of the model averaged over all seasons, is denoted in the lower right corner.}
    \label{fig:crop_mbclmm_pred}
\end{figure}
\newpage
\section*{Table 1: CLMM Results}
\begin{table}[htbp!]
\centering
\caption{Fixed effects Estimates and Random Effect Std.Dev}
\label{table:mbclmm-est}
{\begin{tabular*}{\linewidth}{@{}l@{\extracolsep{\fill}}l@{\extracolsep{\fill}}r@{\extracolsep{\fill}}r
@{\extracolsep{\fill}}r@{\extracolsep{\fill}}r@{\extracolsep{\fill}}r@{\extracolsep{\fill}}c@{}}
\hline
Crop & Parameter & Estimate & SE & Wald z & SD \\
\hline
\multirow{7}{*}{Sorghum} 
& Planted & 6.147 & 0.062 & 98.841 $\star$ & 0.231 \\
& Headed & -6.148 & 0.091 & -67.896 $\star$ & 0.368 \\
& Coloring & -8.352 & 0.104 & -80.513 $\star$ & 0.431 \\
& Mature & -11.123 & 0.098 & -113.479 $\star$ & 0.373 \\
& Harvested & -11.841 & 0.118 & -100.144 $\star$ & 0.431 \\
& Calendar & 4.921 & 0.166 & 29.628 $\star$ & 0.356 \\
& Thermal & 1.197 & 0.282 & 4.245 $\star$ & 0.624 \\
\hline
\multirow{8}{*}{Soybeans} 
& Planted & 6.850 & 0.096 & 71.517 $\star$ & 0.432 \\
& Emerged & -0.445 & 0.045 & -9.859 $\star$ & 0.201 \\
& Blooming & -5.151 & 0.056 & -91.763 $\star$ & 0.251 \\
& Setting Pods & -6.662 & 0.080 & -83.488 $\star$ & 0.359 \\
& Dropping Leaves & -10.715 & 0.080 & -133.836 $\star$ & 0.359 \\
& Harvested & -14.821 & 0.157 & -94.282 $\star$ & 0.709 \\
& Calendar & 5.822 & 0.665 & 8.751 $\star$ & 2.370 \\
& Thermal & 0.657 & 0.650 & 1.010  \,\,\, & 2.351 \\
\hline
\multirow{6}{*}{Winter Wheat} 
& Planted & 16.818 & 0.142 & 118.749 $\star$ & 0.590 \\
& Emerged & -1.172 & 0.100 & -11.686 $\star$ & 0.383 \\
& Headed & -32.999 & 0.224 & -147.096 $\star$ & 0.900 \\
& Harvested & -37.727 & 0.383 & -98.636 $\star$ & 0.724 \\
& Calendar & 10.966 & 2.055 & 5.338 $\star$ & 5.428 \\
& Thermal & 5.410 & 1.060 & 5.102 $\star$ & 2.776 \\
\hline
\multirow{6}{*}{Oats} 
& Planted   & 10.070 & 0.339 & 29.679 $\star$ & 0.657 \\
& Emerged   & -3.614 & 0.351 & -10.287 $\star$ & 0.429 \\
& Headed    & -8.478 & 0.394 & -21.534 $\star$ & 0.981 \\
& Harvested & -12.590 & 0.394 & -31.952 $\star$ & 0.862 \\
& Calendar  & 6.913 & 3.191 & 2.167 $\bullet$ & 6.256 \\
& Thermal   & 0.479 & 2.995 & 0.160 \,\,\, & 5.872 \\
\hline
\multirow{8}{*}{Dry Beans} 
& Planted & 8.544 & 0.352 & 24.256 $\star$ & 0.918 \\
& Emerged & -1.632 & 0.131 & -12.473 $\star$ & 0.289 \\
& Blooming & -7.674 & 0.391 & -19.638 $\star$ & 1.019 \\
& Setting Pods & -8.677 & 0.445 & -19.509 $\star$ & 1.165 \\
& Dropping Leaves & -11.448 & 0.383 & -29.914 $\star$ & 0.996 \\
& Harvested & -13.234 & 0.476 & -27.828 $\star$ & 1.243 \\
& Calendar & 3.526 & 1.492 & 2.363 $\bullet$ & 3.640 \\
& Thermal & 2.232 & 0.921 & 2.423 $\bullet$ & 2.242 \\
\hline
\multirow{6}{*}{Alfalfa} 
& Planted & 4.108 & 0.127 & 32.31 $\star$ & 0.254 \\
& Second Harvest & -2.223 & 0.108 & -20.656 $\star$ & 0.215 \\
& Third Harvest & -4.262 & 0.109 & -39.260 $\star$ & 0.217 \\
& Fourth Harvest & -6.555 & 0.186 & -35.157 $\star$ & 0.372 \\
& Calendar & 0.446 & 0.020 & 22.264 $\star$ & - \\
& Thermal & 2.996 & 0.021 & 145.797 $\star$ & - \\
\hline
\multirow{4}{*}{Millet} 
& Planted & 5.748 & 0.279 & 20.596 $\star$ & 0.552 \\
& Harvested & -9.880 & 0.390 & -25.358 $\star$ & 0.766 \\
& Calendar & 0.860 & 0.059 & 14.522 $\star$ & - \\
& Thermal & 4.002 & 0.079 & 50.615 $\star$ & - \\
\hline
\multicolumn{6}{l}{Significance level: $\bullet < 0.05, \quad \star < 0.001$}
\end{tabular*}}
\bigskip
\hfill
\end{table}
\newpage
\section*{Code:}
The \textit{Ages of Man} project is available on GitHub and can be accessed through https://github.com/agesofman